\documentclass[12pt]{iopart}

\makeatletter
\long\def\@makecaption#1#2{
  \vskip\abovecaptionskip
  \sbox\@tempboxa{#1. #2}
  \ifdim \wd\@tempboxa >\hsize
    #1. #2\par
  \else
    \hbox to\hsize{\hfil #1. #2\hfil}
  \fi
}
\makeatother

\usepackage{cite}

\usepackage{url}

\usepackage{bm,amssymb}
\expandafter\let\csname equation*\endcsname\relax
\expandafter\let\csname endequation*\endcsname\relax

\usepackage{amsmath}

\DeclareMathOperator{\sech}{sech}
\newcommand{\bq}{\bm{q}}
\newcommand{\bx}{\bm{x}}
\newcommand{\bn}{\bm{n}}

\newcommand{\diff}{\mathrm{d}}

\newcommand{\be}{\bm{e}}

\usepackage{graphicx}% Include figure files
\usepackage{epsfig}
\usepackage{xcolor}
{} % End formatting

\usepackage{xspace}
\newcommand{\Peclet}{P\'eclet\xspace}

\begin{document}

\title[Splitting probabilities of active particles]{Splitting probabilities of confined chiral active Brownian particles}

\author{Sarafa A. Iyaniwura$^1$ and Zhiwei Peng$^{2,\footnotemark[1]}$ }

\address{$^1$Vaccine and Infectious Disease Division, Fred Hutchinson Cancer Center, Seattle, Washington 98109, USA}
\address{$^2$Department of Chemical and Materials Engineering,
University of Alberta, Edmonton, Alberta T6G 1H9, Canada}

\footnotetext[1]{Author to whom any correspondence should be addressed.}

\ead{iyaniwura@aims.ac.za \& zhiwei.peng@ualberta.ca}

\vspace{10pt}
\begin{indented}
\item[]%\today 
\end{indented}

\begin{abstract}
Active particles exhibit self-propulsion, leading to transport behavior that differs fundamentally from passive Brownian motion. In confined or structured domains, activity strongly influence escape probabilities and first-passage behavior. Understanding these effects is essential for describing transport in biological microenvironments, microfluidic devices, and heterogeneous media. In this work, leveraging the backward Fokker--Planck equation, we investigate the splitting probability of chiral active Brownian particles in confined domains, focusing on both a one-dimensional interval and a two-dimensional corrugated channel. Analytical solutions are derived for the one-dimensional case in various asymptotic regimes. In corrugated channels with small aspect ratios, we develop a Fick--Jacobs reduction that yields effective transport equations along the axial direction, whereas for finite aspect ratios, the splitting dynamics are characterized numerically.
We demonstrate how channel geometry, particle activity, and chirality modulate the likelihood of escape through different boundaries. Our results provide quantitative predictions for the transport of active matter in complex environments and highlight the interplay between confinement and activity.
\end{abstract}

\section{Introduction}

The splitting probability is a fundamental concept in stochastic transport theory, quantifying the likelihood that a particle or trajectory exits a domain through a given boundary before reaching another. It naturally arises in first-passage processes and provides a probabilistic framework for characterizing competing escape pathways in confined systems. Because many biological and physical processes involve stochastic motion in bounded domains, splitting probabilities have become a widely used analytical tool in physics, biology, and engineering \cite{rogers2010first, redner2001guide, Klinger2022Splitting, Calvani2023Splitting}.

In cellular and molecular biology, splitting probabilities describe competing molecular fates, such as whether a signaling molecule reaches its target receptor before being degraded or sequestered. In intracellular transport, they quantify the likelihood that diffusing or actively transported cargo reaches specific organelles. In immune system modeling, first-passage and escape probabilities govern search and encounter processes that underlie cellular activation and molecular binding. In fluid mechanics and soft matter physics, splitting probabilities characterize transport through pores, channels, and heterogeneous media, where geometric confinement and drift determine preferential escape routes \cite{holcman2015stochastic, delgado2015conditional}.
More broadly, splitting probabilities provide a unifying framework that connects diffusion theory, reaction–transport processes, and boundary-value problems for stochastic motion in complex environments \cite{redner2001guide,berg2025random,holcman2015stochastic}, making them an essential tool for understanding stochastic dynamics across disciplines.

Recent advances in nonequilibrium statistical physics have driven growing interest in splitting probabilities for active matter systems. Unlike passive Brownian particles, active particles are powered by internal or environmental energy sources, resulting in self-propulsion and directional persistence that fundamentally break detailed balance and render equilibrium statistical physics insufficient. In confined or heterogeneous environments, activity strongly alters escape statistics, induces rectification, and gives rise to geometry-dependent transport phenomena that have no passive counterpart \cite{Bechinger2016Active, angelani2024optimal, peng2024macrotransport, Iyaniwura_2025}.
Chiral active particles introduce an additional layer of complexity through intrinsic rotational dynamics, leading to curved trajectories and orientation-dependent interactions with boundaries \cite{caprini2024self, chan2024chiral}. Chirality can qualitatively modify first-passage and escape behavior by competing with translational persistence, thereby reshaping transport pathways and exit probabilities in confined domains \cite{upadhyaya2024narrow, angelani2024optimal}. These effects are especially relevant in biological contexts, including cell migration, immune surveillance and search strategies, and intracellular transport in crowded or structured environments, where active and chiral motility mechanisms are ubiquitous \cite{wang2022cell, yamamoto2025epithelial, wang2024chemical}.

As a result, analytical and computational studies of first-passage and splitting probabilities have become essential tools for understanding how nonequilibrium motility mechanisms influence target finding, escape, and reaction outcomes in complex geometries. Within mathematical biology, stochastic transport and first-passage frameworks have been widely used to model immune cell signaling, receptor activation, and spatial search processes, including diffusion-controlled reactions and signaling dynamics in confined domains \cite{delgado2015conditional, coombs2009diffusion, holcman2015stochastic}. In parallel, recent asymptotic and numerical studies of escape problems in active and confined systems have clarified how geometry, persistence, and rotational dynamics jointly shape first-passage behavior in both biological and physical settings \cite{angelani2014first,angelani2015run,paoluzzi2020narrow,Iyaniwura_2025, angelani2024optimal,olsen2020escape, caprini2021correlated,baouche2025first,gueneau2024optimal}.

Other stochastic systems studied in the context of active matter include run-and-tumble particles (RTPs), which alternate between ballistic motion and stochastic reorientation. Malakar et al. \cite{malakar2018steady} developed an analytical framework to characterize the steady-state, relaxation, and first-passage properties of an RTP with Brownian diffusion in one dimension. By formulating coupled transport equations and solving them using Laplace transform and spectral methods, they obtained exact results for probability distributions and dynamical observables under various boundary conditions, revealing non-Boltzmann steady states, boundary accumulation, and nontrivial relaxation dynamics driven by persistence. Singh et al. \cite{singh2020run} extended this framework to inhomogeneous media, where particle velocity and tumbling rates vary spatially. Their analysis showed that such heterogeneity leads to non-uniform density profiles, with accumulation in regions of low velocity or tumbling rate, and significantly alters transport and first-passage behavior.

Boundary effects in active particle systems were investigated recently by Gu\'{e}neau and Touzo \cite{gueneau2024relating}, who analyzed the relationship between absorbing and hard-wall boundary conditions by incorporating an external potential into the run-and-tumble framework. They established a connection between these boundary conditions and demonstrated how confinement can be effectively described through modified dynamics, providing a unified perspective on boundary-induced effects in non-equilibrium systems and highlighting how different boundary conditions influence steady-state distributions and transport properties. Building on this, Baouche et al. \cite{baouche2026spatiotemporal} extended the Siegmund duality framework to active Brownian particles confined between two walls, deriving analytical predictions for first-passage times and spatial distributions across low- and high-activity regimes. They further demonstrated that ABPs with absorbing and hard-wall boundary conditions are Siegmund duals, and showed that the stationary distribution between hard walls relaxes to a wall-accumulated state given by the derivative of the splitting probability.

Despite these advances, theoretical understanding of splitting probabilities for active and chiral particles remains limited, especially in spatially structured or confined geometries such as corrugated channels. While that of passive Brownian particles in bounded domains is relatively well understood \cite{coombs2009diffusion, grebenkov2026competition, holcman2015stochastic}, the interplay of self-propulsion, rotational dynamics, and geometric confinement presents significant analytical challenges \cite{peng2024macrotransport, Bechinger2016Active}. Developing predictive theoretical and computational frameworks for such systems is critical for describing transport in biological microenvironments, engineered microfluidic devices, and heterogeneous porous media \cite{Bechinger2016Active, Alonso2019Transport, caprini2021correlated}. A recent study by Malgaretti et al.\ \cite{malgaretti2023splitting} investigated the splitting probabilities of passive and active particles using the Fick--Jacobs approach. In their model, an active particle possesses a binary orientational degree of freedom, meaning it can move only to the left or to the right. The swim velocity flips sign randomly at a constant rate.

In this work, we investigate the splitting probability of both chiral and achiral active Brownian particles within confined domains. The motion of a chiral active Brownian particle (CABP) is governed by the following Langevin equations \cite{sevilla2016diffusion,liebchen2022chiral,caprini2023chiral,chan2024chiral}:
\begin{equation}
\label{eq:Langevin}
\begin{split}
\frac{\diff \bx}{\diff t} &= U_s\,  \bq(\theta) + \sqrt{2D_x}\,\boldsymbol{\eta}_x(t), \\[1ex]
\frac{\diff \theta}{\diff t} &= \Omega + \sqrt{2D_\theta}\,\eta_{\theta}(t),
\end{split}
\end{equation}
where $U_s$ is the self-propulsion speed of the particle and $\bq = \cos \theta\; \be_x + \sin \theta \;\be_y$ denotes the orientation vector; $\be_x$ and $\be_y$ are the unit basis vectors in the $x$ and $y$ directions, respectively. In Eq.~\eqref{eq:Langevin}, $\Omega$ is the intrinsic angular velocity characterizing chirality. The coefficients $D_x$ and $D_\theta$ are the translational and rotational diffusion constants, respectively. The stochastic processes $\boldsymbol{\eta}_x(t)$ and $\eta_{\theta}(t)$ represent independent Gaussian white noise in the translational and rotational degrees of freedom, respectively. The dynamics of a standard active Brownian particle (ABP) are recovered by setting $\Omega \equiv 0$. Starting from the Langevin dynamics in Eq.~\eqref{eq:Langevin}, we derive the corresponding backward Fokker--Planck equation for the splitting probability in \ref{app:Derivation-PDE}. In general, we show that the splitting probability  satisfies 
\begin{equation}
    U_s \bq \cdot \nabla p_R +  D_x\nabla^2  p_R + \Omega \, \frac{\partial p_R}{\partial \theta}  +D_\theta\, \frac{\partial^2p_R}{\partial\theta^2} = 0,
\end{equation}
where $p_R$ denotes the splitting probability, defined as the probability that the particle escapes the domain through the right exit rather than the left. Consequently, $p_R$ is unity for a particle starting at the right exit and vanishes for one starting at the left. Additionally, reflecting boundary conditions must be imposed at the impermeable walls to ensure zero net flux.

Applying this backward Fokker--Planck framework, we analyze the splitting behavior of these active particles across two specific confinement scenarios: a one-dimensional (1D) interval (\S \ref{Sec:1D_Chiral}) and a two-dimensional (2D) corrugated channel (\S \ref{Sec:2D}). For the 1D system, we develop asymptotic solutions in both the weak- and strong-activity regimes. For the 2D channel, we first construct a Fick--Jacobs approximation for the splitting probability, showing that it provides reasonable estimates even for finite-aspect-ratio channels. We then compute the splitting probabilities by solving the full backward Fokker–Planck equation numerically. Overall, our results provide both analytical and numerical insights into how active motion and confinement interact to control splitting behavior.

\section{Active particles in a one-dimensional interval}\label{Sec:1D_Chiral}

\begin{figure}
\centering 
\includegraphics[width=3.5in]{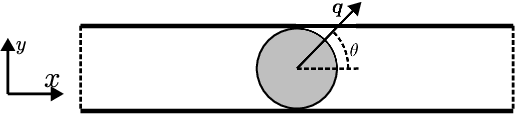}
\caption{Schematic illustration of an active particle in a narrow slit geometry.  The orientation vector, $\bq$, forms an angle $\theta$ with the horizontal ($x$) axis. The particle's translational motion is restricted to the horizontal direction while its orientation angle remains free to rotate continuously in 2D.}
\label{fig:schematic_1d}
\end{figure}

Consider a chiral active Brownian particle (CABP)  in a 1D interval $[-L,L]$, with absorbing boundaries at both ends, as illustrated in Fig.~\ref{fig:schematic_1d}. This setup models a CABP in a narrow slit geometry, where translational motion is restricted to the $x$-axis, while its orientation angle remains free to rotate continuously. Let $p_R$ be the splitting probability that the CABP escapes the interval through the right boundary before the left, given that it started from position $x \in [-L,L]$ with orientation $\bq$. In \ref{app:Derivation-PDE}, we show that $p_R$ satisfies
\begin{equation}\label{eq:SplitProb_PDE_1D}
\begin{split}
U_s \bq \cdot \nabla p_R +  D_x \nabla^2 p_R &+ \Omega \frac{\partial p_R}{\partial \theta }   +D_\theta \frac{\partial^2 p_R}{\partial \theta^2} = 0, \quad  x \in [-L,L], \quad \theta \in [0,2\pi]; \\[1ex]
        p_R(x=-L, \bq) = 0; & \quad p_R(x=L, \bq) = 1,
\end{split}
\end{equation}
where $U_s$ is the particle's swim speed, $\bq$ is the swimming direction ($\bq\cdot\bq=1$), $D_x$ is the translational diffusivity, $\Omega$ denotes  its chirality (i.e., angular velocity), and $D_\theta$ is the rotational diffusivity. Here, $\nabla = \frac{\partial }{\partial \bx}$ is the translational gradient operator.

To write Eq.~\eqref{eq:SplitProb_PDE_1D} in dimensionless form, we scale length by $L$ and time with diffusive timescale $\tau_D = L^2/D_x$. The dimensionless form of Eq.~\eqref{eq:SplitProb_PDE_1D} is given by 
\begin{equation}\label{eq:Non_Dim_SplitProb_1D}
 \begin{split}
   Pe \cos\theta\; &  \frac{\partial p_R}{\partial x} +  \, \frac{\partial^2 p_R}{\partial x^2}  + \chi\;  \frac{\partial p_R}{\partial \theta} + \gamma\; \frac{\partial ^2p_R}{\partial \theta^2}=0, \quad  x \in [-1,1], \quad \theta \in [0,2\pi]; \\[1ex]
        p_R(-1, \theta) &= 0,  \quad p_R(1, \theta) = 1; \qquad p_R(x, 0) = p_R(x, 2\pi),
\end{split}
\end{equation}
where 
\begin{equation}
\label{eq:dimensionless-parameters}
    Pe = U_sL/D_x,   \quad \chi = \Omega L^2/D_x, \quad \text{and}\quad  \gamma =L^2D_\theta/D_x.
\end{equation}
Here, the \Peclet number $(Pe)$ measures the activity of the particles, $\chi$ is the dimensionless chirality, and $\gamma$ characterizes the relative strength of rotational diffusion compared to translational diffusion. For $Pe = 0$, after integrating out the orientational degree of freedom,  the equations reduce to those describing passive Brownian particles (PBPs). When $\chi = 0$, they reduce to the equations for achiral active Brownian particles (ABPs). To obtain the splitting probability for active particles with an initially uniform orientation distribution, we average over all possible orientations:
\begin{equation}
\label{eq:pR-average}
    \langle p_R \rangle (x)  = \frac{1}{2\pi}\int_0^{2\pi} p_R(x, \theta) \; \mathrm{d}\theta \; .
\end{equation}
Consequently,  $\langle p_R\rangle$  depends only on the starting position $x$.

Next, we analyze the splitting probability in the weak-activity limit ($Pe \ll 1$). Consider the asymptotic expansion of $p_R$ in the regime of $Pe \ll 1$, given by
\begin{equation}\label{Eq:Perturb_Expand}
    p_R(x, \theta) = p_0(x, \theta) + Pe \; p_1(x, \theta) +Pe^2\; p_2(x, \theta)+ O\left( Pe^3 \right). 
\end{equation}
Substituting the asymptotic expansion in Eq.~\eqref{Eq:Perturb_Expand} into Eq.~\eqref{eq:Non_Dim_SplitProb_1D}, at leading-order, we obtain the PDE
\begin{equation}\label{eq:SP_LeadOrder}
 \begin{split}
  \frac{\partial^2 p_0}{\partial x^2} & + \chi\;  \frac{\partial p_0}{\partial \theta} + \gamma\; \frac{\partial ^2p_0}{\partial \theta^2}=0 \, ; \\[1ex]
        p_0(-1, \theta) = 0, & \quad p_0(1, \theta) = 1; \qquad p_0(x, 0) = p_0(x, 2\pi).
\end{split}
\end{equation}
At this order, activity is absent, and the solution for Eq.~\eqref{eq:SP_LeadOrder} is 
\begin{equation}
\label{eq:passive-solution}
    p_0(x) = (x+1)/2.
\end{equation} This corresponds to the splitting probability of a (passive) Brownian particle. 

The PDEs  for higher-order terms in the asymptotic expansion \eqref{Eq:Perturb_Expand} have the same structure and can be written as 
\begin{equation}\label{eq:SP_HigherOrder}
 \begin{split}
  \frac{\partial^2 p_n}{\partial x^2}  + \chi\;  \frac{\partial p_n}{\partial \theta} + \gamma\; &\frac{\partial ^2p_n}{\partial \theta^2}= -\cos(\theta) \frac{\partial p_{n-1}}{\partial x}, \quad n \geq 1  \, ; \\[1ex]
        p_n(\pm 1, \theta) = 0, & \quad p_n(x, 0) = p_n(x, 2\pi),
\end{split}
\end{equation}
where $n$ represents the order of the PDE in the asymptotic expansion. To solve these PDEs, we write their solutions in terms of the Green's function $G(x, \theta; \xi, \phi)$ as follows
\begin{equation}\label{eq:SP_HigherOrder_SOl}
 \begin{split}
p_n(x,\theta) = -\int_{-1}^{1} \int_{0}^{2\pi} G(x, \theta; \xi, \phi) \, \cos(\phi) \, \partial_{\xi} p_{n-1} (\xi,\phi) \; \diff \phi \; \diff  \xi , \quad n \geq 1,
\end{split}
\end{equation}
where the Green's function satisfies
\begin{equation}\label{eq:GreensFuncPDE}
 \begin{split}
  \frac{\partial^2 G}{\partial x^2}  + \chi\;  \frac{\partial G}{\partial \theta} + \gamma\; &\frac{\partial ^2 G}{\partial \theta^2}= \delta(x-\xi)\, \delta(\theta - \phi)\, ; \\[1ex]
        G(\pm 1, \theta; \xi, \phi) = 0, & \quad G(x, 0; \xi, \phi) = G(x, 2\pi; \xi, \phi).
\end{split}
\end{equation}
Since $G$ is periodic in $\theta$, we can write the solution of the Green's function in Fourier mode as 
\begin{equation}\label{eq:GreensFunc}
 \begin{split}
  G(x, \theta; \xi, \phi) = \frac{1}{2\pi} \sum_{k =-\infty}^{\infty } g_k(x,\xi) e^{ik(\theta - \phi)},
\end{split}
\end{equation}
where for each mode, $g_k(x,\xi)$ satisfies the following ordinary differential equation (ODE)
\begin{equation}\label{eq:g_ODE}
 \begin{split}
 \frac{\diff^2 g_k}{\diff x^2} + \lambda_k \,g_k  =  \delta(x-\xi), \quad  g_k(\pm1, \xi) = 0,  \qquad k \in \mathbb{Z},
\end{split}
\end{equation}
where $\lambda_k = i \chi k - \gamma k^2$, and $i = \sqrt{-1}$ is the imaginary unit. Observe that this is a forced Sturm-Liouville type ODE. Using the Green's function method, we obtain 
\begin{equation}
g_k(x,\xi)
=
\frac{1}{\sqrt{\lambda_k} \sinh(2\sqrt{\lambda_k})}
\begin{cases}
\sinh\!\Big(\sqrt{\lambda_k} (x+1)\Big)\,
\sinh\!\Big(\sqrt{\lambda_k} (1-\xi)\Big),
& x < \xi, \\[6pt]
\sinh\!\Big(\sqrt{\lambda_k} (\xi+1)\Big)\,
\sinh\!\Big(\sqrt{\lambda_k} (1-x)\Big),
& x > \xi.
\end{cases}
\end{equation}

At $O(Pe)$, we may write the solution as $p_1 = A_1(x) \cos\theta + B_1(x)\sin\theta$, where
\begin{equation}
\label{eq:A1}
    A_1(x)= - c\; 
\mathrm{Re} \Big[ (\gamma - i \chi) \cosh(\alpha^*) \cosh(x \alpha) - \gamma \cosh(\alpha^*) \cosh(\alpha) \Big],
\end{equation}
\begin{equation}
\label{eq:B1}
B_1(x) = \frac{1}{2}c \;
\mathrm{Re} \Big[ (-i \gamma + \chi) \cosh(\alpha^*) \cosh(x \alpha) - \chi \cosh(\alpha^*) \cosh(\alpha) \Big]. 
\end{equation}
In Eqs. \eqref{eq:A1} and \eqref{eq:B1},  $\mathrm{Re}$ denotes the real part of a complex quantity,  $\alpha = \sqrt{\gamma + i \chi} = a + i b$, $\alpha^*$ is the complex conjugate of $\alpha$, and 
\begin{equation}
    c=\frac{ e^{2 a} \, \bigl| -1 + \tanh(\alpha) \bigr|^2 }{ 2 (\gamma^2 + \chi^2) }.
\end{equation}
We note that, $p_1$ does not contribute to $\langle p_R\rangle$ due to symmetry. Using Eqs. \eqref{eq:A1} and \eqref{eq:B1}, one can obtain the solution of $p_2$. Writing $p_2 =C_0(x) + A_2(x)\cos(2\theta) + B_2(x)\sin(2\theta)$, we obtain 
\begin{equation}\label{Eq:C0}
    C_0(x) = \frac{1}{4}\mathrm{Re}\left[ \frac{\sech(\sqrt{\gamma+i\chi})\sinh(x\sqrt{\gamma+i\chi})-x \tanh(   \sqrt{\gamma+i\chi})}{\left(\gamma+i\chi\right)^{3/2}}\right].
\end{equation}
With this, the splitting probability for particles with an initially uniform orientation distribution can be expanded as
\begin{equation}\label{Eq:_SplitProb-two-term}
    \langle  p_R \rangle = p_0 + Pe^2 \; C_0 +o(Pe^2),
\end{equation}
where $o(Pe^2)$ denotes high-order terms. Therefore, in the weak-activity limit, the leading correction to the passive result ($p_0$) appears at $O(Pe^2)$. For achiral ABPs ($\chi=0$), $C_0$ simplifies to 
\begin{equation}
    C_0(x) = \frac{\sech \left(\sqrt{\gamma }\right) \sinh \left(\sqrt{\gamma } \;x\right)-x \tanh \left(\sqrt{\gamma }\right)}{4 \gamma ^{3/2}}.
\end{equation}

\begin{figure}
\centering 
\includegraphics[width=3in]{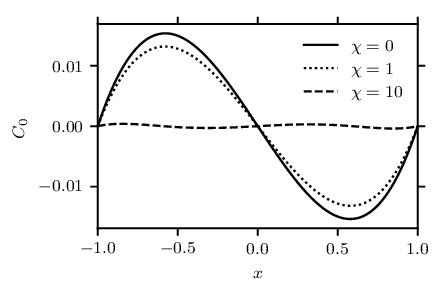}
\caption{Plots of $C_0(x)$, defined in Eq.~\eqref{Eq:C0}, as a function of the initial position of the particle $x$ for different values of chirality ($\chi$). In all cases, $\gamma=0.1$.}
\label{fig:splitting_orderPe2}
\end{figure}
In Fig.~\ref{fig:splitting_orderPe2}, we plot $C_0(x)$, defined in Eq.~\eqref{Eq:C0}, as a function of $x$ for different values of chirality $(\chi)$. For ABPs ($\chi=0$), $C_0$ oscillates around the midpoint of the interval and is positive for $x<0$ and negative for $x>0$. At the midpoint ($x=0$), symmetry requires $\langle p_R \rangle = 1/2$, and consequently  $C_0=0$. For particles starting at $0<x<1$, activity slightly reduces the probability that they escape through the right exit compared to the passive case, so that $C_0 <0$. Unlike the passive case, where the splitting probability is determined solely by diffusion and the distance to the exit, active particles can reduce the bias toward the right exit through their swimming motion, which allows them to move toward the left. Conversely, for particles starting at $-1<x<0$, their bias towards the left exit ($p_L$) is reduced,  and consequently $p_R=1-p_L$ is enhanced in this region.

\begin{figure}
\centering 
\includegraphics[width=3in]{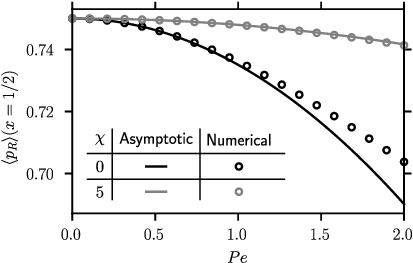}
\caption{Comparison between the two-term asymptotic solution given in Eq.~\eqref{Eq:_SplitProb-two-term}  and the full numerical solution of Eq.~\eqref{eq:Non_Dim_SplitProb_1D}  for the splitting probability $\langle p_R\rangle$ evaluated at  $x=1/2$, for $\chi = 0$ and $\chi=5$. The two-term asymptotic approximation at $x=1/2$ is given by $3/4+Pe^2 C_0(1/2)$. In all cases, $\gamma=0.1$. } 
\label{fig:small-Pe-compare}
\end{figure}

As $\chi$ increases, the magnitude of the $O(Pe^2)$ correction gradually decreases, since chirality effectively reduces the persistence of swimming motion.  In the limit $\chi \to \infty$,  $C_0(x) \to 0$, so that $\langle p_R \rangle = p_0$. Consequently, the splitting probability approaches that of a passive Brownian particle, consistent with the fact that the motion of CABPs becomes effectively diffusive in the large-chirality regime.

For arbitrary $Pe$, we solve Eq.~\eqref{eq:Non_Dim_SplitProb_1D} numerically using the finite element method (FEM) implemented in \texttt{FreeFem++} \cite{hecht2012new}. In Fig.~\ref{fig:small-Pe-compare}, we compare the two-term asymptotic solution given in Eq.~\eqref{Eq:_SplitProb-two-term} with the full numerical solution at $x=1/2$ for $\chi=0$ and $\chi=5$. For $Pe \lesssim 1$, excellent agreement is observed for both cases. For ABPs ($\chi=0$), the small-$Pe$ asymptotic solution begins to deviate from the full solution as $Pe$ exceeds $\approx 1$. In contrast, for CABPs, chirality effectively reduces particle activity; consequently, the two-term asymptotic solution remains in very good agreement even at $Pe=2$.

\begin{figure}
\centering 
\includegraphics[width=5in]{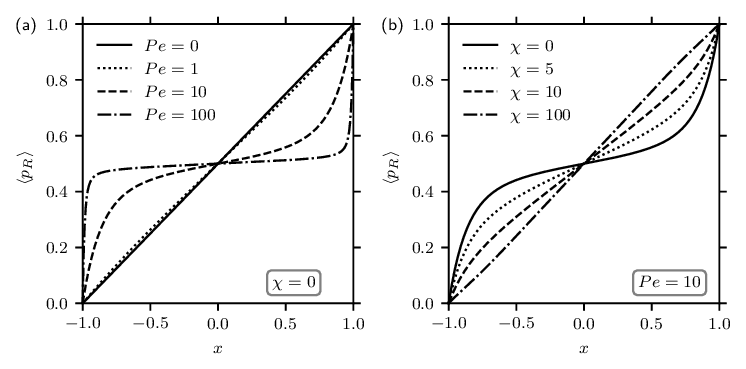}
\caption{Plots of the splitting probability $\langle p_R\rangle$ as a function of the particle’s initial position $x$. 
(a) $\langle p_R\rangle$ for ABPs ($\chi=0$) at different values of $Pe$. 
(b) $\langle p_R\rangle$ for CABPs at different values of chirality $\chi$ ($Pe=10$). 
In all cases, $\gamma=0.1$.}
\label{fig:splitting_1d}
\end{figure}

The splitting probability $\langle p_R\rangle$, computed using FEM, is shown in Fig.~\ref{fig:splitting_1d} as a function of the initial position $x$. Fig.~\ref{fig:splitting_1d}(a) presents $\langle p_R\rangle$ for ABPs ($\chi=0$) at $Pe=0,\,1,\,10,$ and $100$.
For $Pe=0$, the splitting probability reduces to that of a passive Brownian particle and varies linearly with the initial position [see Eq.~\eqref{eq:passive-solution}], increasing monotonically from the left boundary to the right boundary. As the activity increases ($Pe>0$), deviations from this linear behavior emerge, reflecting the influence of self-propulsion. Specifically, for particles starting at $-1<x<0$, activity enhances the probability of exiting through the right boundary, whereas for particles starting at $0<x<1$, activity reduces $\langle p_R\rangle$ relative to the passive case. This behavior is consistent with the weak-activity analysis (see Fig.~\ref{fig:splitting_orderPe2}).

With further increase in $Pe$, the dependence of $\langle p_R\rangle$ on the initial position weakens. In the large-$Pe$ limit, the splitting probability becomes nearly uniform across the domain interior and saturates at $\langle p_R\rangle \approx 0.5$, indicating an equal likelihood of escape through either boundary. At large $Pe$, the escape dynamics are governed primarily by self-propulsion. With a uniform initial orientation, particles starting in the bulk are equally likely to escape through either the left or right exit. In the limit $Pe \to \infty$, boundary layers develop near $x=\pm 1$, where $\langle p_R\rangle$ rapidly adjusts to satisfy the different boundary conditions.

Fig.~\ref{fig:splitting_1d}(b) shows the splitting probability $\langle p_R\rangle$ for a fixed activity level $Pe=10$ and varying chirality, illustrating the effect of chirality on escape dynamics. Consistent with the behavior of $C_0(x)$ shown in Fig.~\ref{fig:splitting_orderPe2}, increasing chirality progressively suppresses the influence of activity on the splitting probability. Physically, chirality shifts the curve towards the linear profile.  Starting from the achiral case $\chi=0$, increasing $\chi$ leads to a reduction of $\langle p_R\rangle$ for particles starting at $x<0$ and an enhancement for particles starting at $x>0$. In the limit $\chi \to \infty$, the splitting probability becomes indistinguishable from that of PBPs, reflecting the effective loss of directional persistence as rapid rotational dynamics render the particle motion diffusive.

To characterize the boundary-layer structure observed in Fig.~\ref{fig:splitting_1d}(a), we consider a singular perturbation in the high-activity regime ($Pe \gg 1$). 
Defining $\epsilon = 1/Pe$, we rewrite Eq.~\eqref{eq:Non_Dim_SplitProb_1D} as
\begin{equation}
    \cos\theta\,  \frac{\partial p_R}{\partial x}+\epsilon \, \frac{\partial^2 p_R}{\partial x^2} +\epsilon \, \chi\, \frac{\partial p_R}{\partial \theta} + \epsilon\, \gamma\, \frac{\partial^2 p_R}{\partial \theta^2}=0. 
\end{equation}
Setting $\epsilon=0$ eliminates the $\theta$-derivatives entirely, implying the formation of boundary layers as $\epsilon \to 0$. In the bulk, the leading-order balance gives $\frac{\partial p_R}{\partial x} =0$, which implies that $p_R=p_R(\theta)$ at leading order. To resolve the boundary layer at $x=1$, we consider a stretched coordinate $s= (1-x)/\epsilon$.  The boundary-layer thickness is dictated by a balance between the swimming term and the diffusive term, $\cos\theta\,  \frac{\partial p_R}{\partial x} \sim  \epsilon \, \frac{\partial^2 p_R}{\partial x^2}$.

In the boundary layer, we expand $p_R$ as $p_R= \tilde{p}_0(s, \theta) + \cdots$, where the leading-order equation is 
\begin{equation}
\label{eq:p0-BL-right}
\begin{split}
    -\cos\theta \, \frac{\partial \tilde{p}_0}{\partial s} + \frac{\partial ^2 \tilde{p}_0 }{\partial s^2}=0; \qquad 
    \tilde{p}_0(0, \theta)=1. 
\end{split}
\end{equation}
Integrating Eq.~\eqref{eq:p0-BL-right}  yields $\tilde{p}_0 = 1 +\left( e^{s\cos\theta} - 1\right)c_1(\theta)$, where $c_1(\theta)$ remains to be determined. This  boundary-layer solution must match the bulk solution as $s \to \infty$. To enforce this matching, we distinguish between the two cases according to the sign of $\cos\theta$. For $\cos\theta >0$, we must set $c_1(\theta) \equiv 0$ since $e^{s\cos\theta}$ diverges as $s\to\infty$. As a result, the bulk solution for this case must be $p_R \equiv 1$ as well. Thus, for $\cos\theta >0$, no boundary layer forms  at $x=1$; instead, the boundary layer is located at the opposite boundary, $x=-1$. 

Conversely, for $\cos\theta <0$, the boundary layer is located at $x=1$. In this case, the bulk solution must satisfy the boundary condition at $x=-1$. Matching the bulk solution with the boundary-layer  solution yields $c_1(\theta) \equiv 1$. As a result, for $\cos\theta <0$,  a composite solution that is valid in the entire domain is given by $p_R = e^{(1-x)\cos\theta/\epsilon}$. Following the same approach, for $\cos\theta >0$, the corresponding composite solution is given by  $ p_R = 1 - e^{-(1+x)\cos\theta/\epsilon}$. Taken together, we obtain
\begin{equation}
\label{eq:pR-theta-largePe}
 p_R(x, \theta) =
    \begin{cases}
       e^{Pe(1-x)\cos\theta}&  \cos\theta <0, \\ 
       1-e^{-Pe(1+x)\cos\theta}& \cos\theta >0,
    \end{cases}
\end{equation}
which is the leading-order asymptotic solution to $p_R$ as $Pe \to \infty$. 

\begin{figure}
\centering 
\includegraphics[width=5.9in]{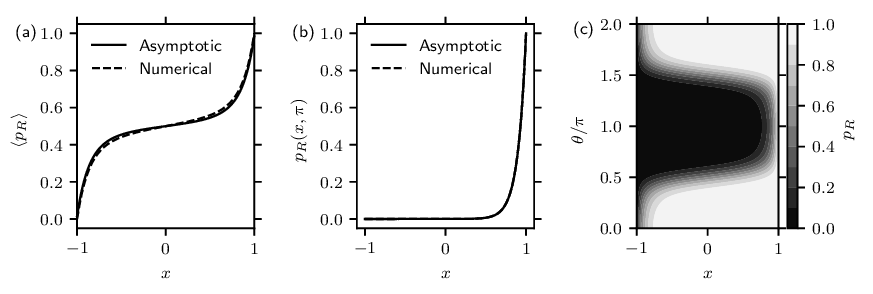}
\caption{ (a) Comparison between the numerical (FEM) and leading-order high-activity asymptotic solutions for $\langle p_R \rangle$ at $Pe=10$. We note that the asymptotic solution [Eq.~\eqref{eq:1D-large-Pe-result}] for $\langle p_R \rangle$ is independent of both $\gamma$ and $\chi$. (b) Comparison between the numerical (FEM) and leading-order high-activity asymptotic solutions for $p_R(x,\pi)$ at $Pe=10$. We note that in panel (b), the numerical and asymptotic [Eq.~\eqref{eq:pR-theta-largePe}] solutions are visually indistinguishable. (c) Contour plot of the numerical solutions for $p_R$ at $Pe=10$. For the numerical results shown, we used $Pe=10$, $\gamma=0.1$,  and $\chi=0$.}
\label{fig:largePe}
\end{figure}

The above boundary-layer analysis does not apply near $\theta=\pi/2$ (or $3\pi/2$), where $\cos\theta$ vanishes and the assumed dominant balance $\cos\theta\,  \frac{\partial p_R}{\partial x} \sim  \epsilon \, \frac{\partial^2 p_R}{\partial x^2}$ breaks down. To resolve the splitting probability near $\theta=\pi/2$, one needs to consider a stretched coordinate in $\theta$ given by $\beta = (\theta-\pi/2)/\delta$, where $\delta \ll 1$. In this region, the dominant balance is given by  $Pe\, \cos\theta\, \frac{\partial p_R}{\partial x} \sim \frac{\partial^2 p_R}{\partial \theta^2}$, which leads to a thickness of $\delta= \epsilon^{1/3}$. Since $p_R \leq O(1)$ throughout the domain, this inner-layer solution does not contribute to the integral in Eq. \eqref{eq:pR-average} at leading order. Therefore, the leading-order splitting probability $\langle p_R\rangle$ is given by 
\begin{equation}
\label{eq:1D-large-Pe-result}
\begin{split}
        \langle p_R \rangle &\sim   \int_0^{\frac{\pi}{2}} \left(1-e^{-Pe(1+x)\cos\theta}\right) \mathrm{d}\theta  + \int_{\frac{\pi}{2}}^{\frac{3\pi}{2}}e^{Pe(1-x)\cos\theta}\mathrm{d}\theta +  \int_{\frac{3\pi}{2}}^{2\pi}\left(1-e^{-Pe(1+x)\cos\theta}\right) \mathrm{d}\theta , \\ 
        &=\frac{1}{2} \Big{[}1-L_0\left(Pe(1-x)\right)+L_0\left(Pe(1+x)\right)+I_0\left(Pe(1-x)\right)-I_0\left(Pe(1+x)\right) \Big{]},
\end{split}
\end{equation}
where $I_0$ is the modified Bessel function of the first kind of order zero, and $L_0$ is the modified Struve function of order zero.

In Figs.~\ref{fig:largePe}(a) and (b), we present the asymptotic solutions for the splitting probability and compare these results with the numerical solutions of the full problem, Eq.~\eqref{eq:Non_Dim_SplitProb_1D}. Fig.~\ref{fig:largePe}(a) compares the FEM solution with the leading-order asymptotic expression for the splitting probability $\langle p_R\rangle$, given in Eq.~\eqref{eq:1D-large-Pe-result}. Excellent agreement is observed throughout the entire domain, validating the leading-order asymptotic approximation in the high-activity regime. Similarly, Fig.~\ref{fig:largePe}(b) compares the FEM solution with the leading-order boundary-layer approximation evaluated at $\theta=\pi$. The numerical and asymptotic solutions are visually indistinguishable.

In Fig.~\ref{fig:largePe}(c), we show a contour plot of the splitting probability $p_R$ as a function of the  initial position and orientation, computed using FEM. In terms of $\theta$, the splitting probability is maximized when the particle initially points toward the target boundary at $x=1$ (i.e., $\theta=0$ or $2\pi$), even when the starting position is closer to the left boundary at $x=-1$. As the initial orientation deviates from this direction, $p_R$ decreases, reaching its minimum when the particle points toward the left boundary. Additionally, for a fixed orientation, the splitting probability increases monotonically as the particle starts closer to the target boundary. The boundary-layer structure is clearly visible in Fig.~\ref{fig:largePe}(c).

\section{Active particles in a corrugated channel\label{Sec:2D}}

\begin{figure}
\centering 
\includegraphics[width=3in]{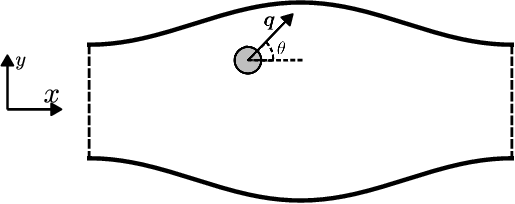}
\caption{Schematic illustration (not to scale) of an active particle in the unit cell of a periodic corrugated channel for $\alpha >0$. When $\alpha < 0$, the narrow neck lies at the midpoint along $x$, while the exits are wide. In the theoretical formulation, the particle is treated as a point. The dashed lines indicate the left and right exits. }
\label{fig:schematic_2d}
\end{figure}

In the previous section, we analyzed the splitting probability for active particles on a one-dimensional interval with flat side walls, where geometry played no explicit role beyond setting the domain size. We now extend this framework to a corrugated channel, in which the width varies periodically along the axial direction. Unlike the uniform interval, such geometric modulation induces spatially varying confinement. To assess how geometric corrugation modifies the escape dynamics, we now consider the splitting probability of active particles in the unit cell of a corrugated channel, where the two exits are located at $x=-L$ and $x=L$, respectively, as illustrated in Fig.~\ref{fig:schematic_2d}.

Let $p_R$ be the splitting probability that the CABP escapes the unit cell through the right boundary before the left, given that it started from position $\bx \equiv (x,y)$ with orientation $\bq$. 
The splitting probability $p_R$ satisfies
\begin{equation}\label{eq:SplitProb_PDE_Channel}
\begin{split}
U_s \bq \cdot \nabla p_R +  D_x\nabla^2 & p_R + \Omega \, \frac{\partial p_R}{\partial \theta}  +D_\theta\, \frac{\partial^2p_R}{\partial\theta^2} = 0; \\[1ex]
        p_R(x=-L, y, \bq) &= 0; \quad p_R(x=L, y, \bq) = 1  \,; \\[1ex]
        \bn \cdot \nabla p_R = 0, \quad & \quad y= \pm H(x), 
\end{split}
\end{equation}
where $H(x) = H_0 \big{(}1 + \alpha \cos(\pi x/L) \big{)}$, $ \lvert  \alpha \rvert < 1$,
defines the upper wall of the channel. The lower wall is at $y=-H(x)$.  In Eq. \eqref{eq:SplitProb_PDE_Channel}, $\bn$ denotes the outward unit normal vector on the walls
$y = H(x)$ and $y = -H(x)$. Because the top and bottom walls are no-flux boundaries, a Neumann boundary condition is imposed \cite{Iyaniwura_2025}.

 We follow a similar approach used to non-dimensionalize the problem in \S \ref{Sec:1D_Chiral} to obtain the dimensionless form of \eqref{eq:SplitProb_PDE_Channel}, given by
\begin{equation}\label{eq:NonDim_SplitPr_CorrugChannel}
 \begin{split}
   Pe \cos\theta\;  \frac{\partial p_R}{\partial x} & \,+ Pe  \sin\theta\;   \frac{\partial p_R}{\partial y} +  \, \frac{\partial^2 p_R}{\partial x^2}  + \frac{\partial^2 p_R}{\partial y^2}  + \chi\;  \frac{\partial p_R}{\partial \theta} + \gamma\; \frac{\partial^2 p_R}{\partial \theta^2}=0; \\[1ex]
      p_R(-1, y, \theta) &= 0,  \quad p_R(1, y, \theta) = 1; \qquad p_R(x,y, 0) = p_R(x,y, 2\pi)\,; \\[1ex]
      -\varepsilon h' \, \frac{\partial p_R}{\partial x} + & \frac{\partial p_R}{\partial y} = 0, \quad y=\varepsilon h(x)\, ; \qquad  \varepsilon h' \, \frac{\partial p_R}{\partial x} + \frac{\partial p_R}{\partial y} = 0, \quad y= - \varepsilon h(x),
\end{split}
\end{equation}
where $h(x) = 1 + \alpha \cos(\pi x)$, $h^\prime = \frac{\mathrm{d} h}{\mathrm{d}x}$,  $\varepsilon = H_0/L$ is the characteristic aspect ratio of the channel. The parameters \(Pe\), \(\chi\), and \(\gamma\) are defined as in Eq. \eqref{eq:dimensionless-parameters}.
Our objective is to use this dimensionless framework to characterize the splitting probability that a CABP introduced into the domain exits through the right opening before escaping through the left opening. In the small-aspect-ratio limit, $\varepsilon \ll 1$, we reduce the governing equations with two spatial dimensions to an effective one-dimensional description in space, while retaining the orientational dynamics,  using the Fick--Jacobs approximation \cite{Jacobs2012-sx,zwanzig1992diffusion,Reguera2001,kalinay2005projection,kalinay2006exact,kalinay2006corrections,jain2023fick,zhao2025active}. We subsequently consider channels with finite aspect ratios through FEM simulations of Eq. \eqref{eq:NonDim_SplitPr_CorrugChannel}.

\subsection{Fick--Jacobs theory}
In the small-aspect-ratio limit, we assume that the transverse dynamics relaxes rapidly to local equilibrium. On this basis, we extend the Fick--Jacobs framework to analyze Eq. \eqref{eq:NonDim_SplitPr_CorrugChannel} by integrating each term across the channel at fixed  $x$. The integral of the spatial Laplacian 
\begin{equation}
\label{eq:Leibniz-laplacian}
    \int_{-\varepsilon h}^{\varepsilon h} \left(  \frac{\partial^2 p_R}{\partial x^2}+\frac{\partial^2 p_R}{\partial y^2}\right) \mathrm{d}y = \frac{\partial}{\partial x}\int_{-\varepsilon h}^{\varepsilon h} \frac{\partial p_R}{\partial x}\mathrm{d}y,
\end{equation}
where we have used the Neumann boundary conditions at $y= \pm \varepsilon h$ together with the  Leibniz integration rule.  We then approximate $p_R$ as 
\begin{equation}
\label{eq:FJ-expansion}
    p_R(x, y,\theta) = \overline{p}_R(x, \theta) + \cdots,
\end{equation}
where the leading approximation is independent of the transverse coordinate $y$ and higher-order corrections are neglected. From Eqs. \eqref{eq:Leibniz-laplacian} and \eqref{eq:FJ-expansion}, we obtain 
\begin{equation}
    \int_{-\varepsilon h}^{\varepsilon h} \left(  \frac{\partial^2 p_R}{\partial x^2}+\frac{\partial^2 p_R}{\partial y^2}\right) \mathrm{d}y \approx\frac{\partial }{\partial x}\left( 2 \varepsilon h \frac{\partial \overline{p}_R}{\partial x}\right). 
\end{equation}
The same approximation can be applied to the integrals of the remaining terms in Eq. \eqref{eq:NonDim_SplitPr_CorrugChannel}. Taken together, we obtain
\begin{equation}\label{eq:FJ-finaleq}
 \begin{split}
    Pe \cos\theta \; &\frac{\partial\overline{p}_R }{\partial x} + \frac{1}{h}\frac{\partial }{\partial x}\left( h \frac{\partial \overline{p}_R}{\partial x}\right) + \chi \frac{\partial \overline{p}_R}{\partial \theta} + \gamma \frac{\partial^2 \overline{p}_R}{\partial \theta^2}=0, \\[1ex]
        \overline{p}_R(-1, \theta) &= 0,  \quad \overline{p}_R(1, \theta) = 1; \qquad \overline{p}_R(x, 0) = \overline{p}_R(x, 2\pi).
\end{split}
\end{equation}
For a flat channel, $h=const.$, Eq. \eqref{eq:FJ-finaleq} reduces to Eq. \eqref{eq:Non_Dim_SplitProb_1D}.

\begin{figure}
\centering 
\includegraphics[width=5.9in]{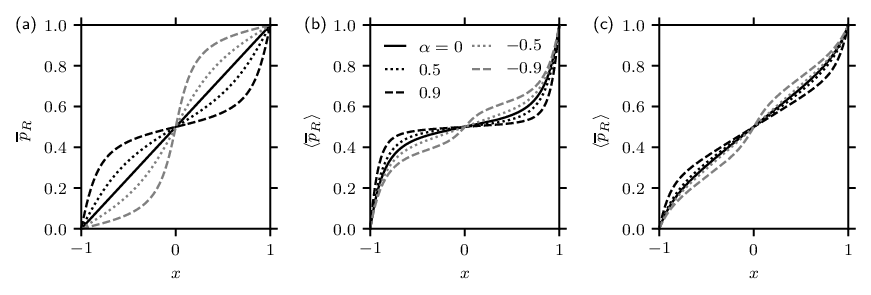}
\caption{ Splitting probabilities computed from the Fick–Jacobs theory. (a) Plots of $\overline{p}_R$ for passive particles ($Pe=0$), defined in Eq.~\eqref{eq:FJ-passive},  as a function of $x$ for different values of $\alpha$. (b) Plots of $\langle \overline{p}_R \rangle $ for ABPs ($Pe=10, \chi=0$) as a function of $x$ for different values of $\alpha$. (c) Plots of $\langle \overline{p}_R \rangle $ for CABPs ($Pe=10, \chi=10$) as a function of $x$ for different values of $\alpha$. In all cases, $\gamma=0.1$. Results  in (b) and (c) are from FEM simulations of Eq. \eqref{eq:FJ-finaleq}.  In all cases, setting $\alpha=0$ reduces the result to that of a one-dimensional interval (see \S \ref{Sec:1D_Chiral}). The legends for panels (a)--(c) are shown in (b).}
\label{fig:FJ}
\end{figure}

In the passive case ($Pe=0$), Eq.~\eqref{eq:FJ-finaleq} can be shown to admit the solution 
\begin{equation}
\label{eq:FJ-passive}
    \overline{p}_R = \frac{1}{\pi} \tan ^{-1}\left(\frac{(1-\alpha ) \tan \left(\frac{\pi  x}{2}\right)}{\sqrt{1-\alpha ^2}}\right) + \frac{1}{2}.
\end{equation}
Furthermore, when $\alpha=0$, Eq. \eqref{eq:FJ-passive} reduces to $\overline{p}_R = (x+1)/2$, as given in Eq. \eqref{eq:passive-solution}. For ABPs and CABPs, we solve Eq. \eqref{eq:FJ-finaleq} numerically using FEM. Following Eq. \eqref{eq:pR-average}, we consider the splitting probability for particles with an initially uniform orientation distribution, $\langle \overline{p}_R\rangle = \int_0^{2\pi} \overline{p}_R/(2\pi) \mathrm{d}\theta$. For passive particles, $\langle \overline{p}_R\rangle =\overline{p}_R $.

In Fig.~\ref{fig:FJ}(a), we show the splitting probability $\overline{p}_R$ for PBPs ($Pe=0$) in a corrugated channel, as defined in Eq.~\eqref{eq:FJ-passive}, for three values of the corrugation amplitude: $\alpha=0$ (no axial modulation; flat channel), $\alpha=0.5$, and $\alpha=0.9$.
For $\alpha=0$, the channel reduces to a straight geometry, and $\overline{p}_R$ coincides with the splitting probability for PBPs in a 1D interval, increasing linearly with the initial position as particles start closer to the right exit. As the corrugation amplitude $\alpha$ increases, geometric confinement progressively biases transport: the splitting probability increases for particles starting at $x<0$ and decreases for those starting at $x>0$, relative to the straight channel case. When $\alpha > 0$, the right exit lies in a narrow section, making it more difficult for particles starting at $0 < x < 1$ to escape through it compared to a flat channel. As a result, $\overline{p}_R$ decreases for $0 <x <1$ as $\alpha$ increases.  On the other hand, for $\alpha <0$, the narrow neck lies at the midpoint of the domain; in this case,  $\overline{p}_R$ for $0<x<1$ increases as $\alpha$ becomes more negative.

Figures~\ref{fig:FJ}(b) and~(c) show the splitting probability $\langle \overline{p}_R\rangle$, computed from FEM simulations of the Fick--Jacobs equation ~\eqref{eq:FJ-finaleq}, for particles initialized with a uniform orientation distribution. Figure~\ref{fig:FJ}(b) presents results for ABPs ($Pe=10,\ \chi=0$), while Fig.~\ref{fig:FJ}(c) shows the corresponding results for CABPs ($Pe=10,\ \chi=10$).
In both cases, the qualitative effect of the corrugation amplitude $\alpha$ on the splitting probability is similar to that observed for PBPs in Fig.~\ref{fig:FJ}(a). Specifically, increasing $\alpha$ enhances the splitting probability for particles starting at $x<0$ and reduces it for those starting at $x>0$, reflecting the increasing influence of geometric confinement and associated entropic barriers, despite the left-right symmetry of the channel.

Comparing Figs.~\ref{fig:FJ}(b) and~(c), we observe that chirality reduces the impact of activity on the splitting probability, in a manner similar to the 1D case. For a fixed $Pe$, increasing chirality effectively reduces the particle persistence. For fixed values of $\alpha$, the enhancement of $\langle \overline{p}_R\rangle$ for particles starting at $x<0$ is less pronounced for CABPs than for ABPs, and similarly, the reduction for particles starting at $x>0$ is also less significant. Together, these results demonstrate how chirality modulates the interplay between activity and geometric confinement, leading to a partial suppression of confinement-induced variations in the splitting probability.

\subsection{ Finite-aspect-ratio  channels}
\begin{figure}
\centering 
\includegraphics[width=5in]{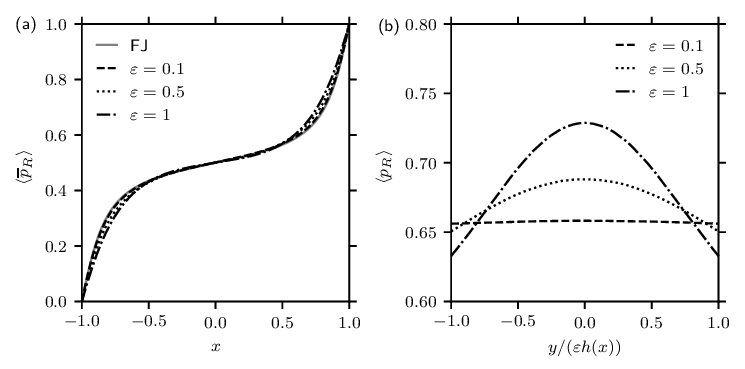}
\caption{ (a) Comparison of the numerical solutions of the Fick--Jacobs (FJ) approximation [Eq. \eqref{eq:FJ-finaleq}] and the full PDE model [Eq. \eqref{eq:NonDim_SplitPr_CorrugChannel}] for the splitting probability $\langle \overline{p}_R \rangle $ of ABPs. (b) Plots of the splitting probability $\langle p_R\rangle$ at $x=0.75$ as a function of the normalized $y$ coordinate, $y/(\varepsilon h)$, from full numerical solutions. The physical parameters are given by $Pe=1, \gamma=0.1, \chi=0$, and $\alpha=0.9$.}
\label{fig:FJvsNum}
\end{figure}

For channels with finite aspect ratios, we solve Eq.~\eqref{eq:NonDim_SplitPr_CorrugChannel} numerically using FEM. To examine the accuracy of the Fick--Jacobs approximation, we compute the cross-sectionally averaged splitting probability as
\begin{equation}
    \langle \overline{p}_R\rangle(x) = \frac{1}{2\pi}\int_0^{2\pi} \mathrm{d}\theta \frac{1}{2\varepsilon h(x)}\int_{-\varepsilon h}^{\varepsilon h}p_R(x, y, \theta)  \mathrm{d}y. 
\end{equation}
In Fig.~\ref{fig:FJvsNum}(a), we compare the Fick--Jacobs approximation with the full numerical solution for $\langle \overline{p}_R\rangle(x)$. For small aspect ratios (e.g., $\varepsilon=0.1$), the  Fick--Jacobs approximation agrees very well with the full numerical solution across the entire domain in $x$. As the aspect ratio increases, the discrepancy between the Fick–Jacobs approximation and the full solution gradually increases. Overall, even at $\varepsilon = 1$, the Fick--Jacobs solution continues to provide a good qualitative approximation.

\begin{figure}
\centering 
\includegraphics[width=5.9in]{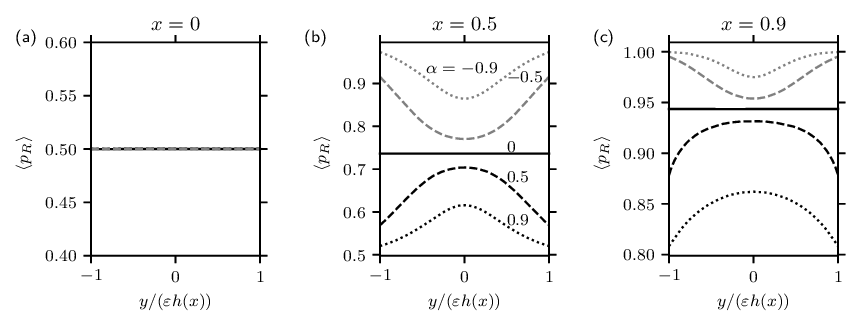}
\caption{Plots of the splitting probability $\langle p_R\rangle$ as a function of the normalized $y$ coordinate $y/(\varepsilon h)$ at different values of the axial coordinate: (a) $x=0$, (b) $x=0.5$, and (c) $x=0.9$, computed numerically from the full PDE model [Eq.~\eqref{eq:NonDim_SplitPr_CorrugChannel}]. The physical parameters are given by $Pe=1, \gamma=0.1, \chi=0$, and $\varepsilon=1$. In (a), all curves collapse onto a single line; the legend for the different lines is shown in (b).}
\label{fig:finite}
\end{figure}

Quantitatively, the Fick--Jacobs approximation systematically underestimates the splitting probability for particles starting at $0<x<1$ because it assumes instantaneous transverse equilibration. As shown in Fig.~\ref{fig:FJvsNum}(b), this assumption is valid for small $\varepsilon$, where $\langle p_R\rangle$ has a  nearly uniform profile across the transverse coordinate $y/(\varepsilon h(x))$.  As $\varepsilon$ increases,  $\langle p_R\rangle$ becomes increasingly nonuniform across the transverse direction. For finite aspect ratios, the maximum of $\langle p_R\rangle$ at a fixed $x$ is located at the center of the channel, $y=0$. In this regime, active particles do not fully equilibrate across the channel. Their persistent self-propulsion, together with interactions with the walls, biases motion toward the right exit. These effects are not captured by the Fick--Jacobs reduction.

In Fig.~\ref{fig:finite}, we show $\langle p_R \rangle$ versus the normalized transverse coordinate $y/(\varepsilon h(x))$ for various $x$ positions and values of $\alpha$, further illustrating the splitting dynamics of ABPs in finite-aspect-ratio channels. For a flat channel [$\alpha=0$; solid lines in Figs.~\ref{fig:finite}(a)-(c)], the splitting probability, $\langle p_R \rangle$, is independent of the transverse coordinate and increases as the starting position moves closer to the right exit [Fig.~\ref{fig:finite}(a), $x=0$; Fig.~\ref{fig:finite}(b), $x=0.5$; Fig.~\ref{fig:finite}(c), $x=0.9$ ]. The increase of $\langle p_R \rangle$ with the starting position $x$ is preserved for finite channel modulations (i.e., $\alpha = 0.5, 0.9$). For particles starting at the midpoint ($x=0$), $\langle p_R \rangle = 1/2$ and is independent of the channel modulation [Fig.~\ref{fig:finite}(a)]. As the starting position moves closer to the right exit [Fig.~\ref{fig:finite}(b)-(c)], $\langle p_R \rangle $ becomes nonuniform across the channel.

We now compare the cases $\alpha > 0$ and $\alpha < 0$. For $\alpha > 0$, the corrugated channel is narrow at the exits (see Fig.~\ref{fig:schematic_2d}), whereas for $\alpha < 0$, the narrow neck is located at the midpoint between the two exits. At a fixed $x$ location, for example $x=0.5$ [Fig.~\ref{fig:finite}(b)] or $x=0.9$ [Fig.~\ref{fig:finite}(c)], the splitting probability attains its maximum at the center ($y=0$) of the channel when $\alpha >0$. In contrast, for $\alpha <0$, the maximum occurs at the walls, while the center corresponds to the minimum. For ABPs that are stuck at the wall, they tend to slide along the boundary towards the ``easy'' direction. In the case $\alpha <0$, this motion directs them towards the wider region, namely the right exit. Consequently, the splitting probability at the wall is higher than compared to in the bulk of the channel. For $\alpha > 0$, the opposite behavior occurs, as the particles now tend to slide towards the horizontal center of the channel.

\section{Discussion}

In this work, we investigated the splitting probability of confined active particles, using the backward Fokker--Planck equation. Our analysis considered two canonical geometries: a one-dimensional interval and a two-dimensional corrugated channel. In both cases, we quantified how particle activity, chirality, and confinement influence the  probability that the particles escapes the domain through the right exit before the left.

For the one-dimensional interval, we derived analytical expressions for the splitting probability in both the weak- and high-activity regimes. These asymptotic results elucidate how self-propulsion and orientational persistence systematically bias escape relative to passive diffusion, producing activity-dependent deviations from the classical linear splitting profile. In the weak-activity limit, the behavior approaches that of passive Brownian motion, with perturbative corrections arising from finite persistence. In contrast, in the high-activity regime, the dynamics become increasingly dominated by directed swimming. In this regime, the splitting probability becomes independent of the initial position and converges to $1/2$, except for particles starting sufficiently close to the boundaries, where the probability approaches the imposed boundary conditions. This gives rise to boundary layers near both ends of the interval. These boundary layers were systematically characterized using singular perturbation theory. Beyond providing physical insight into the mechanisms governing escape in the simplest confined setting, these analytical solutions serve as useful limiting benchmarks for more complex geometries. In particular, they establish reference behaviors against which the effects of geometric confinement, entropic barriers, and effective dimensional reduction in corrugated channels can be assessed.

In corrugated channels, we applied a Fick--Jacobs reduction to analyze transport in small-aspect-ratio geometries and employed numerical simulations to study channels with finite aspect ratios. We found that the Fick--Jacobs approximation provides quantitatively accurate predictions for small aspect ratios but progressively deviates from the full solution as the channel aspect ratio increases. Moreover, our results reveal that the transverse profile of the splitting probability becomes increasingly nonuniform with higher aspect ratios, typically exhibiting maxima near the channel center. The introduction of chirality and channel modulation further alters these profiles, either suppressing or enhancing the likelihood of escape through a specific exit.

Our analysis focuses on idealized channels and does not account for hydrodynamic interactions or crowding effects, which can be significant in dense or complex biological environments. Future extensions could explore the influence of time-dependent boundary conditions, interactions between multiple active particles, and heterogeneous activity profiles, providing a more comprehensive understanding of transport in realistic microenvironments. 
Another natural extension of this work is the study of splitting probabilities for active particles in two-dimensional domains with partially reactive boundaries, through which particles can escape, similar to the analysis performed for Brownian particles in \cite{grebenkov2026competition}. Such an approach would allow the investigation of how boundary reactivity, particle activity, and geometric confinement collectively influence escape dynamics.

In our model, both translational and rotational noises are present.  To isolate the effect of translational noise, it is of interest to consider purely active motion, where $D_x \equiv 0$. This limit eliminates passive diffusion and allows one to focus on transport driven solely by self-propulsion and rotational fluctuations. While the backward Fokker--Planck equation still applies, analytical progress may become more challenging. We note that the forward Fokker--Planck equation for purely active motion has been analyzed in different contexts \cite{lee2013active,ezhilan2015distribution,dulaney2020waves}. Alternatively, one can employ the Langevin equation to characterize the dynamics of purely active motion \cite{peng2020upstream}.

Overall, our study highlights the roles of confinement geometry, activity, and chirality in shaping the first-passage statistics of active particles. We show that self-propulsion fundamentally modifies escape dynamics relative to passive diffusion. Our results underscore the inherently non-equilibrium nature of active transport and are consistent with previous studies demonstrating that first-passage properties of active particles differ qualitatively from those of passive Brownian motion due to persistence, anisotropy, and boundary interactions \cite{Bechinger2016Active, malakar2018steady, Woillez2019Activated, iyaniwura2024asymptotic, malgaretti2023splitting}.

More broadly, our results contribute to the growing theoretical framework for first-passage phenomena in active matter, where propulsion, rotational dynamics, and confinement jointly determine escape statistics and transport efficiency. While first-passage problems in bounded domains are classical for passive diffusion, their extension to active systems remains fundamentally challenging because persistent motion and orientational degrees of freedom generate intrinsically non-equilibrium transport mechanisms. 

The backward Fokker--Planck approach adopted here provides a systematic and predictive framework for quantifying these effects across different geometries. Beyond the specific configurations considered, our findings suggest general principles governing how activity and confinement control transport. These insights may inform the design of synthetic microswimmer systems and guide the interpretation of active transport in microfluidic devices, biological microenvironments, and other structured or heterogeneous media.

\appendix

\section{Derivation of the backward Fokker--Planck equation\label{app:Derivation-PDE}}

We derive the Fokker--Planck equation governing the splitting probability of a chiral active Brownian particle (CABP) confined to a bounded domain, quantifying the likelihood of escape through a specified exit, starting from an underlying random-walk description of the particle dynamics \cite{Iyaniwura_2025}.

Let $p_R(\bx,\theta)$ denote the splitting probability for a CABP that starts at position $\bx$ with orientation $\theta$ at time $t$. Consider a small time increment $\Delta t$, during which the particle undergoes a stochastic displacement, moving to the new state $(\bx + \Delta \bx,\theta + \Delta \theta)$ at time $t+\Delta t$. Averaging over all possible displacements during this interval, the splitting probability satisfies the backward equation
\begin{equation}\label{Eq:PR_Expectation}
p_R(\bx,\theta)
= \mathbb{E}\!\left[\, p_R(\bx + \Delta \bx,\theta + \Delta \theta) \,\right],
\end{equation}
where $\mathbb{E}[\cdot]$ denotes the expectation over the stochastic increments $\Delta \bx$ and $\Delta \theta$. Taylor expanding $p_R(\bx + \Delta \bx,\theta + \Delta \theta)$ about $(\bx,\theta)$ yields
\begin{equation}\label{Eq:TaylorSeries}
\begin{split}
        p_R(\bx + \Delta \bx,\theta + \Delta \theta) &= p_R(\bx, \theta) + \frac{\partial p_R}{\partial x_i} \Delta x_i +  \frac{\partial p_R}{\partial \theta} \Delta \theta \\ 
        &\quad + \frac{1}{2} \frac{\partial^2 p_R}{\partial x_i \partial x_j} \Delta x_i \Delta x_j 
    +  \frac{1}{2} \frac{\partial^2 p_R}{\partial \theta^2} (\Delta \theta)^2 + \cdots
\end{split}
\end{equation}

Consider the Langevin equations describing the motion of a CABP,
\begin{equation}
\begin{split}
\frac{\diff \bx}{\diff t} &= U_s \bq(\theta) + \sqrt{2D_x}\,\boldsymbol{\eta}_x(t), \\[1ex]
\frac{\diff \theta}{\diff t} &= \Omega + \sqrt{2D_\theta}\,\eta_{\theta}(t),
\end{split}
\end{equation}
where $U_s$ is the self-propulsion speed of the particle and $\bq(\theta)=\cos\theta\,\be_x+\sin\theta\,\be_y$ denotes the orientation. Here, $\Omega$ is the intrinsic angular velocity characterizing chirality. The coefficients $D_x$ and $D_\theta$ are the translational and rotational diffusion constants, respectively. The stochastic processes $\boldsymbol{\eta}_x(t)$ and $\eta_{\theta}(t)$ represent independent Gaussian white noise in the translational and rotational directions, respectively.

Over a small time interval $\Delta t$, the corresponding discrete-time updates using Euler--Maruyama scheme are
\begin{equation}\label{Eq:DiscreteLangevin}
\begin{split}
\Delta \bx &= U_s \bq(\theta)\,\Delta t + \sqrt{2D_x\,\Delta t}\,\boldsymbol{W}_x, \\[1ex]
\Delta \theta &= \Omega\,\Delta t + \sqrt{2D_\theta\,\Delta t}\,W_{\theta},
\end{split}
\end{equation}
where $\boldsymbol{W}_x$ and $W_{\theta}$ are independent standard Gaussian random variables with zero mean and unit variance in the translational and rotational directions, respectively. Using the discrete Langevin increments defined in Eq.~\eqref{Eq:DiscreteLangevin}, the first and second moments of the stochastic increments are
\begin{equation}\label{Eq:DiscreteExpect}
\begin{split}
\mathbb{E}[\Delta x_i] &= U_s \bq\, \Delta t; \quad \mathbb{E}[\Delta \theta] = \Omega \, \Delta t; \\[1ex]
\mathbb{E}[\Delta x_i \, \Delta x_j] = U_s^2 q_i  q_j &(\Delta t)^2 + 2 D_x \,  \Delta t \, \delta_{ij} ; \quad 
\mathbb{E}[(\Delta \theta)^2] = \Omega^2 (\Delta t)^2 + 2 D_\theta \, \Delta t  ,
\end{split}
\end{equation}
where $\delta_{ij}$ denotes the Kronecker delta. In deriving these expressions, we have used the statistical properties of the Gaussian increments, $
\mathbb{E}[\boldsymbol{W}_x]=\boldsymbol{0}, \mathbb{E}[W_{\theta}]=0, 
\mathbb{E}[W_{x,i}W_{x,j}]=\delta_{ij}, \mathbb{E}[W_{\theta}^2]=1$.

Now, substituting the expansion in Eq.~\eqref{Eq:TaylorSeries} into the expectation in \eqref{Eq:PR_Expectation}, we have
\begin{equation}\label{Eq:Expectation2}
\begin{split}
\mathbb{E}\!\left[\, p_R(\bx + \Delta \bx,\theta + \Delta \theta) \,\right] &= p_R(\bx, \theta) +\frac{\partial p_R}{\partial x_i} U_s q_i \Delta t \\[1ex]
& \quad + \frac{1}{2} \frac{\partial^2 p_R}{\partial x_i \partial x_j} \Big{(} U_s^2 q_i  q_j (\Delta t)^2 + 2 D_x \,  \Delta t \, \delta_{ij}  \Big{)}\\[1ex]
& \quad + \frac{\partial p_R}{\partial \theta} \Omega \Delta t
+ \frac{1}{2}\frac{\partial^2 p_R}{\partial \theta^2} \Big{(} \Omega^2 (\Delta t)^2 + 2 D_\theta \, \Delta t \Big{)} + \cdots
\end{split}
\end{equation}
Substituting Eq.~\eqref{Eq:Expectation2} into \eqref{Eq:PR_Expectation} and simplifying, we have
\begin{equation}\label{Eq:DiffEquation}
\begin{split}
    \Delta p_R & = \frac{\partial p_R}{\partial x_i} U_s q_i \Delta t + \frac{1}{2} \frac{\partial^2 p_R}{\partial x_i \partial x_j} \Big{(} U_s^2 q_i  q_j (\Delta t)^2 + 2 D_x \,  \Delta t \, \delta_{ij}  \Big{)}\\[1ex]
& \quad +  \frac{\partial p_R}{\partial \theta} \Omega \Delta t
+ \frac{1}{2}\frac{\partial^2 p_R}{\partial \theta^2} \Big{(} \Omega^2 (\Delta t)^2 + 2 D_\theta \, \Delta t \Big{)} + \cdots
\end{split}
\end{equation}
where $\Delta p_R = \mathbb{E}\!\left[\, p_R(\bx + \Delta \bx,\theta + \Delta \theta) \,\right] - p_R(\bx, \theta)$. Dividing through by $\Delta t$ and taking the limit as $\Delta t \to 0$, we obtain the backward Fokker--Planck equation [see also Eq.~\eqref{eq:SplitProb_PDE_1D}] as
\begin{equation}
\begin{split}
U_s \bq \cdot \nabla p_R +  D_x \nabla^2 p_R &+ \Omega \frac{\partial p_R}{\partial \theta }   +D_\theta \frac{\partial^2 p_R}{\partial \theta^2} = 0.
\end{split}
\end{equation}

\section{Finite element simulations}
Let $H^1(\mathcal{D})$ be the standard Sobolev space of functions with square-integrable first derivatives, where  $\mathcal{D} = [-1, 1] \times [0, 2\pi]$. We define the trial space $V$ for the solution $p_R$, which must satisfy the Dirichlet and periodic boundary conditions. The corresponding test space $V_0$ consists of functions $v$ that are periodic in $\theta$. We multiply the governing equation [Eq.~\eqref{eq:Non_Dim_SplitProb_1D}] by an arbitrary test function $v \in V_0$ and integrate over the entire domain $\mathcal{D}$. After applying integration by parts, we obtain the weak formulation as
\begin{equation}
\label{eq:weak_form}
\begin{split}
& \iint_{\mathcal{D}} \left( Pe \cos\theta\,\frac{\partial p_R}{\partial x} v - \frac{\partial p_R}{\partial x} \frac{\partial v}{\partial x} + \chi \frac{\partial p_R}{\partial \theta}v - \gamma \frac{\partial p_R}{\partial \theta}\frac{\partial v}{\partial \theta}\right) \mathrm{d}x \, \mathrm{d}\theta \\ 
&+ \int_0^{2\pi} \frac{\partial p_R}{\partial x} v \Bigl \lvert_{x=-1}^{x=1}\mathrm{d}\theta = 0.
\end{split}
\end{equation}

Following an analogous procedure, we can derive the corresponding weak formulation for $p_R$ in the corrugated channel. The weak formulation is implemented and solved in \texttt{FreeFem++} \cite{hecht2012new}.

\section*{Data availability statement}
The data and code that support the findings of this study are openly available at
\url{https://doi.org/10.5683/SP3/WSJJPE}

\ack 
Z. P. was supported by the Natural Sciences and Engineering Research Council of Canada (NSERC)  under Grant No. RGPIN-2025-05310.

\section*{References}  % Manually insert heading
\bibliographystyle{iopart-num}  % or iopart-num-long or iopart-num-short
\bibliography{references}             % refs.bib is your .bib file

\end{document}